\newcommand{\COMMENT}[1]{}

\newcommand{\reff}[1]{}           

\newcommand{\spaceline}[1]{\vspace{9pt}}
\newcommand{\spacex}[1]{\vspace{1cm}}
\newcommand{\spacexx}[1]{\vspace{2cm}}

\newcommand{\kl}[1]{\left\langle} 
\newcommand{\kr}[1]{\right\rangle} 

\documentclass[10pt,a4paper]{article}  
\usepackage{graphicx}
\usepackage{amssymb}
\usepackage{amsmath}

\usepackage[utf8]{inputenc}  
\usepackage[T1]{fontenc}

\usepackage{here}  

\usepackage[colorlinks]{hyperref}

\def\DREIECK#1{{\def\bull{}%
\count1=0
\loop
\edef\bull{$\bullet$\bull}
\ifnum\count1<#1
\advance\count1 by 1
\centerline{\bull}
\vskip-7.7pt
\repeat
\vskip 7.7pt\relax}}

\begin{document}
\setlength\parindent{0pt}
\setlength{\parskip}{6pt}  
\setlength{\overfullrule}{10pt}

\begin{center}
\Large
\textbf{Touching Loop Patterns with Cellular Automata}
~~\\
~~\\
\normalsize
\textit{Rolf Hoffmann}
\footnotesize
~~\\
Technical University Darmstadt, Germany\\
{hoffmann@informatik.tu-darmstadt.de}\\
\end{center}

\begin{abstract}
\noindent
The objective is the design of a Cellular Automata rule that can form patterns with 
``touching'' loops.
A loop is defined as a closed path of 1-cells  in a 2D grid on a zero background and with a zero border.
A path cell is connected with two of its adjacent neighbors. 
In \textit{touching loops} 
a path cell is also allowed to touch another on a diagonal.  
A CA rule was designed that can evolve stable touching loop patterns.
The rule tries to cover the 2D space by overlapping tiles.
The rule uses so-called \textit{templates}, 5 x 5 matching patterns which are systematically derived from
the given set of 3 x 3 tiles. 
The rule checks the pattern being evolved against a list of templates.
If the outer neighbors of a template match, then the cell's state is set to the template's center value.
Noise is injected if there is no matching template, 
or the tiles are not properly assembled.
Thereby the evolution is driven to the desired loop patterns.
\end{abstract}
\normalsize
%
%
\newpage
\tableofcontents
\newpage

\section{Introduction} 
\label{Introduction} 

This article is based on the presentation given at ACRI,
the 16th International Conference and School “Cellular Automata for Research and Industry” 
September 9-11, 2024,  University of Florence, Italy.

Loop structures are of interest with regard to their emergence, construction principle, functionality, and various other properties. 
Therefore they are a topic of research in many disciplines. 
The challenge here was to generate certain ``large'' loops by local operations only,
namely by a Cellular Automata (CA) rule.

In prior work 
\cite{2014-Hoffmann-Agent-PathPattern} 
the pattern evolution was controlled by finite state machines 
of moving agents that were trained with a Genetic Algorithm. This technique yields good results but needs a lot of 
training effort and is not easy to manage for a variable grid size. 
Therefore we are now evolving patterns by CA rules which are directly constructed using problem specific overlapping tiles 
\cite{2022-2019-Arxiv-Forming-Point-Patterns-by-a-Probabilistic-Cellular-Automata-Rule,2021-Hoffmann-DD-FS-DominoSquareDiamond},
and which are not sensitive against the grid size, the boundary or even obstacles. 
A first CA rule that can evolve loop pattern under cyclic boundary conditions was presented in  
\cite{2023-Hoffmann-Loop},
and more elaborated rules under fixed boundary conditions are presented in
\cite{2024-Hoffmann-Bialecki-Loop-AdvancesInCA}.
The rule designed here is more simple, can evolve loops that touch each other and is applied under fixed boundary conditions. 
The rule uses two types of overlapping tiles,
tiles that are used to build straight path segments 
($\substack{~0~
          \\111
          \\~0~}$),
and
corner tiles 
($\substack{~0~
          \\011
          \\~10}$).
Templates (local matching patterns) are systematically derived  from the tiles.
The CA rule tests the templates against the pattern being evolved. 
If there is a match in the outer neighborhood of a $5 \times 5$ window,
the cell's state is adjusted to the templates center value, otherwise noise is injected.
In addition the \textit{path condition} has to be fulfilled, i.e. 
each path cell needs three tiles in sequence
that overlap with cover level 3.  

\begin{figure}[H] 
\centering
(a)\includegraphics[width=2.7cm,height=2.7cm]{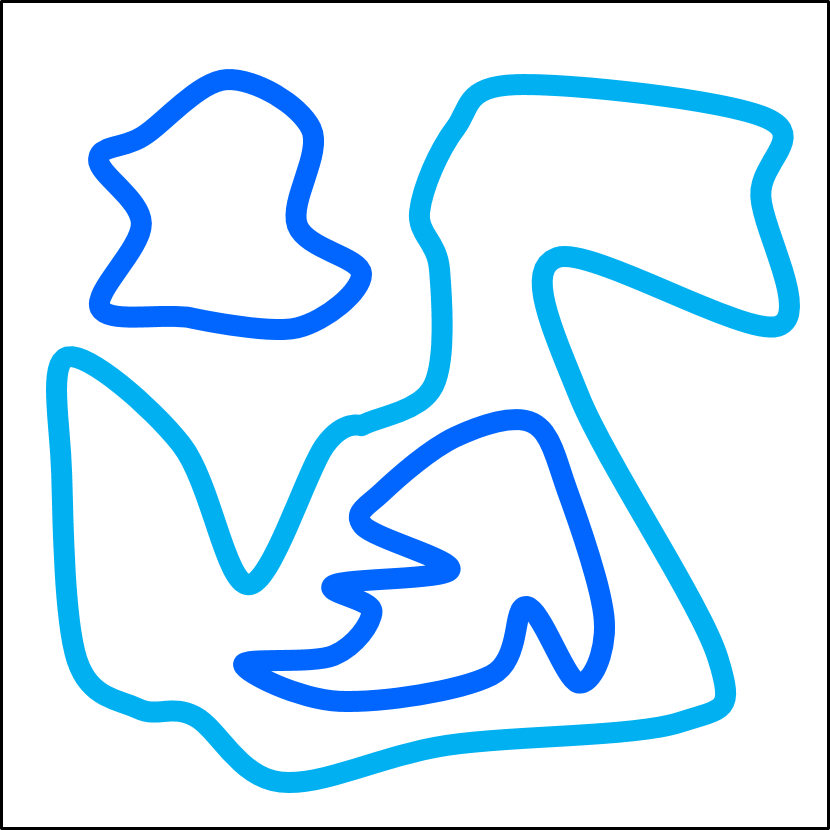}
(b)\includegraphics[width=4cm]{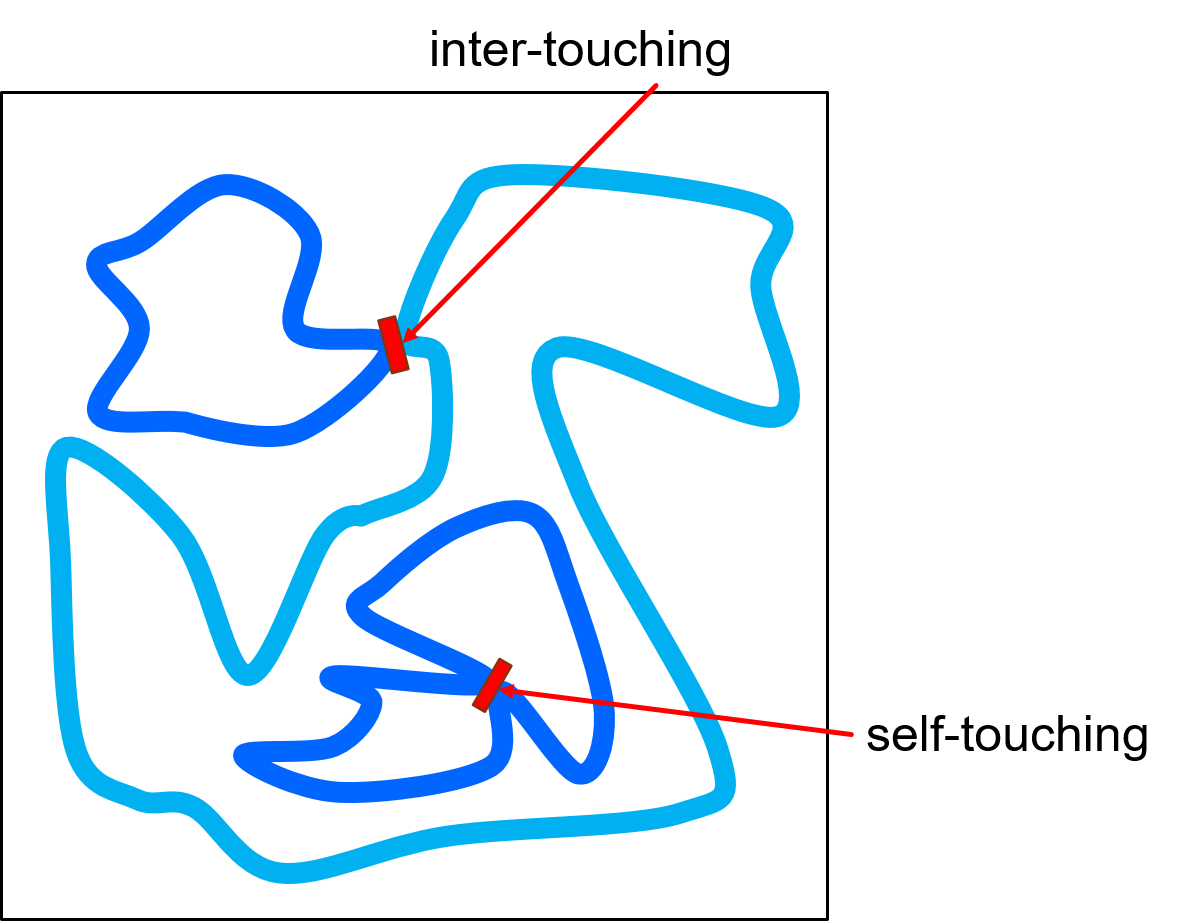}
\caption{
(a) Non-touching loops.
(b) Touching loops.
}
\label{loop-analog1}
\end{figure}

Two different loops can touch each other (\textit{inter-touching} between loops),
and a loop can be \textit{self-touching} when path cells of the same loop touch each other 
(Fig.~\ref{loop-analog1}b).

%
%

\begin{figure}[H] 
\centering
(a)\includegraphics[width=2.7cm]{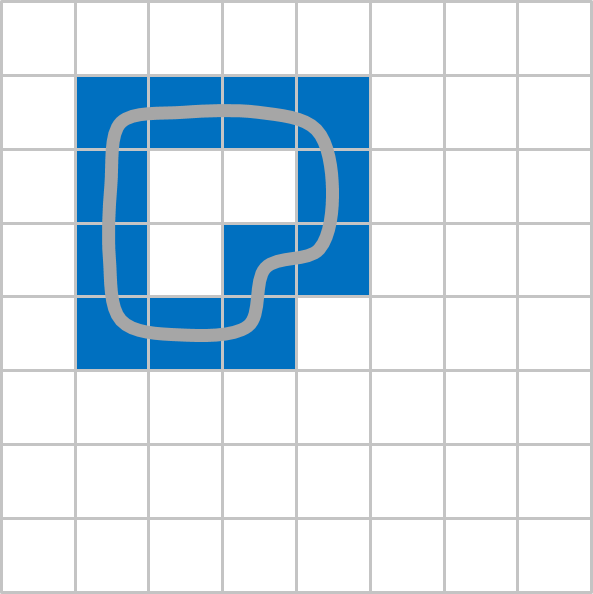}
(b)\includegraphics[width=2.7cm]{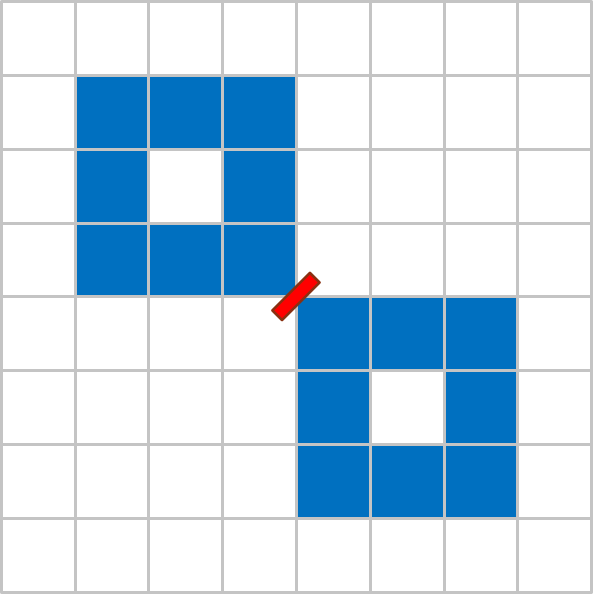}
(c)\includegraphics[width=2.7cm]{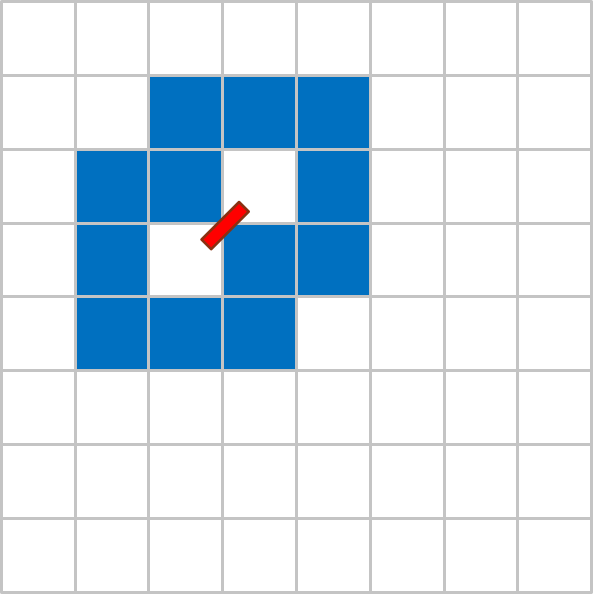}
\caption{
(a) Example of a loop modeled as a closed path of cells in a 2D grid.
(b) Two loops (the most simple ones, two \textit{mini-loops}) are inter-touching each other.
(c) A self-touching loop. 
}
\label{loopGridModel}
\end{figure}

The loops shown in Fig.~\ref{loop-analog1} are ``real'' continuous loops, whilst we will use here discretized loops
fitting into a 2D grid. 
Fig.~\ref{loopGridModel}a shows how a loop is modeled as a cyclic closed path of cells,
where each path cell is connected to exactly two of its adjacent neighbors (NESW, the von-Neumann nearest neighbors: 
\textit{\textbf{N}orth, \textbf{E}ast, \textbf{S}outh, \textbf{W}est}).
Formally the following condition is true for all path cells of a loop:

$(s=1)\wedge ((s^{N} + s^{E}+ s^{S}+ s^{W}) = 2)$

where the cell's main state is $s=1$ for path cells and $s=0$ for the surrounding cells.
The term $s^{N} + s^{E}+ s^{S}+ s^{W}$
gives the number of NESW-neighbors which are in state $s=1$. 

We allow path cells to touch each other on a diagonal, i.e. either two path cells (\textit{NorthEast, SouthWest}) or 
(\textit{NorthWest, SouthEast}) are neighbors in the $3 \times 3$ Moore-neighborhood.   
Path cells which belong to a corner and which lie on a diagonal are not considered as touching. So touching path
cells are given by the local patterns\textit{}
\begin{verbatim}
1 0       or      0 1
0 1               1 0 .
\end{verbatim}
Fig.~\ref{loopGridModel}b shows two \textit{inter-touching} loops (the most smallest ones, called \textit{mini-loops}),
and Fig.~\ref{loopGridModel}c shows a \textit{self-touching} loop. 
%
%

The Problem Statement is

GIVEN

\begin{itemize}
	\item 
  $(n+1) \times (n+1)$ field of cells including a fixed zero-boundary.
  \item
   $n \times n$ inner area of active cells. 
  \item
  The main cell state/color is $s \in \{0,1\}$. 
  Path cells shall have the value 1 and are colored in blue. 
  The value 0 shall be used for the border and the background.
  \item
  The initial configuration is random.
  
\end{itemize}

FIND
\begin{itemize}
	\item 
  A Cellular Automata Rule that can form stable patterns with loops which may inter- or self-touch.  
\end{itemize}

\section{Design basics: Tiles and templates}
\label{Design basics: Tiles and templates}

The rule was designed on the basis of a method already applied successfully 
for evolving point patterns
\cite{2022-2019-Arxiv-Forming-Point-Patterns-by-a-Probabilistic-Cellular-Automata-Rule},
domino patterns
\cite{2021-Hoffmann-DD-FS-DominoSquareDiamond}
\cite{Hoffmann:Deserable-Seredynski-pact-2021b-Minimal-Covering-of-the-Space-by-Domino-Tiles}
sensor networks
\cite{Hoffmann:Deserable-Seredynski-2022-NatCom-Cellular-automata-rules-solving-the-wireless-sensor-network-coverage-problem},
and loop patterns under periodic 
\cite{2023-Hoffmann-Loop}
and fixed 
\cite{2024-Hoffmann-Bialecki-Loop-AdvancesInCA}
boundary conditions. 

\subsection{The method}
\label{The method}

\begin{enumerate}
	\item 
  A \textit{set of tiles }has to be defined that can induce the desired pattern.
	A \textit{tile} is a small pattern consisting of elementary tiles (square cells of minimal size according to the grid)
	which are called \textit{pixels} in order to distinguish them from the \textit{cells} of the aimed pattern.
	The defined tiles (and replicates of them) are then used to fill the given space. 
	Pixels from different tiles are allowed to overlap if there is no conflict in color. 
	If $K$ pixels of the same color are overlaid, then the cover level $v$ of this cell becomes $v=K$
	Depending on the set of tiles and the given space it is not always possible to cover the whole space 
  ($v>0$ for all cells) if that is required.
  \item
  So-called \textit{templates} are derived from the tiles. 
  For each pixel of a tile a template is derived by shifting the tile in such a way that
  the chosen pixel becomes the center. This means that for every tile pixel a specific template is generated.
 \item
  The set of all templates is inserted into a general probabilistic CA rule,
  and the rule is applied in the following way:  
  
  \begin{itemize}
    \item  
    All outer templates (the templates without the center) are tested against the actual cell's neighborhood.  
    If there are hits, they are registered together with their templates.
    \item 
    If there is only one hit, the cell's value is adjusted to the center value of the hitting template.
    If there are several hits, then one of the hitting templates is chosen at random for adjustment.
    \item 
    If there is no hit, then noise is injected. This means that the cell's state is changed into another state
    with a certain probability. In addition, noise can be injected if certain constraints are not fulfilled. 
   \end{itemize}
\end{enumerate}

\subsection{The set of tiles}
\label{SectSetOfTiles}

\begin{figure}[H] 
\centering
(a)\includegraphics[width=2cm]{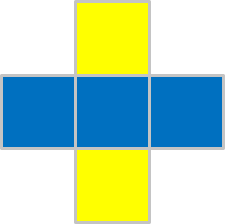}
(b)\includegraphics[width=2cm]{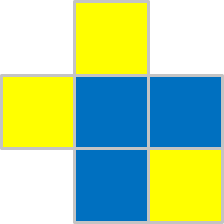}
\caption{
The basic tiles: (a) straight path tile,
(b) corner tile.
The symmetric ones have to be added. 
Blue: pixel value = 1, yellow: pixel value = 0.
}
\label{BasicTiles}
\end{figure}

\begin{figure}[H] 
\centering
\includegraphics[width=12cm]{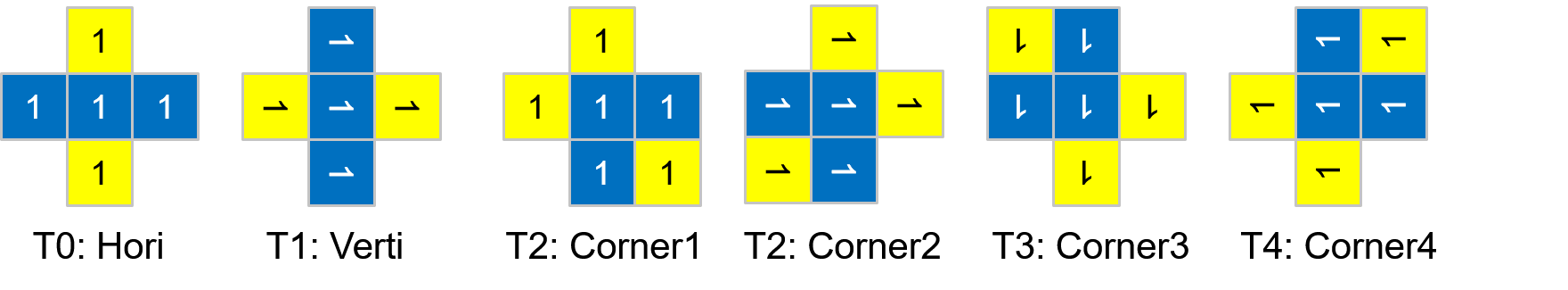}
\caption{
The whole set of tiles, including the symmetric ones.
The numbers indicate the cover/overlay level of the tile pixels, '1' if they are not overlaid.
}
\label{AllTiles}
\end{figure}

For the construction of touching loops only two basic tiles are necessary:
(a) A tile for a straight continuation of the path, and (b) a corner 
(Fig.~\ref{BasicTiles}).
Each tile contains three adjacent one-pixels. 
In order to yield the full set of tiles the symmetric ones have to be added 
(Fig.~\ref{AllTiles}).

%
%
%

In the former publication 
\cite{2024-Hoffmann-Bialecki-Loop-AdvancesInCA} 
 non-touching loops were already found by a slightly different corner tile.
The former and new corner tile differ only in one pixel at the outer
diagonal of the corner point.
\begin{verbatim}
new tile for              former tile for
touching loops            non-touching loops
     0                        0 0  
   0 1 1                      0 1 1
     1 0                        1 0
\end{verbatim}

\subsection{Example of assembling a loop by overlaying tiles}
\label{Example of assembling a loop by overlaying tiles}

\begin{figure}[H] 
\centering
\includegraphics[width=7cm]{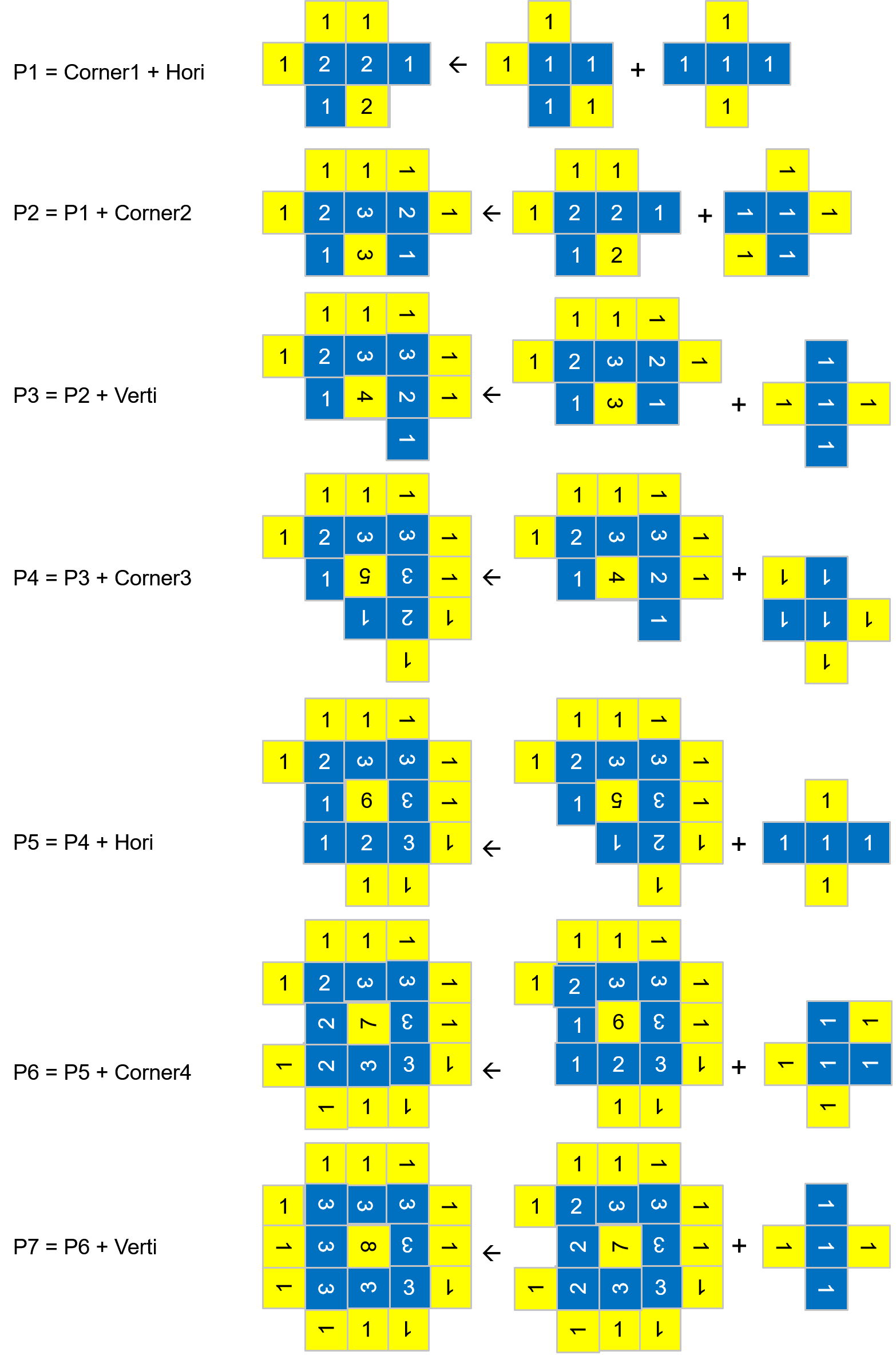}
\caption{
Assembling a $3\times3$ mini-loop step by step.
}
\label{MakeLoop}
\end{figure}

We demonstrate how the simplest possible loop (the \textit{mini-loop}) can be assembled by tiles
(Fig.~\ref{MakeLoop}).
We start with the tile Corner1 and ``add'' (+) the tile Hori,
so we get the pattern P1.
Then we add Corner2 and get P2, and so on.
The add operator joins two patterns where the patterns can be taken
from the set of tiles or already formed patterns. 
When forming the loop, the sequence of joining partial patterns can be 
different, it is not fixed. 
We should notice that the cover level of the loop path cells increase from $v=1$ to $v=3$,
and for the center cell from $v=1$ to $v=8$.

Here we do not give a formal definition of the add-operator,
but the idea is to join two tiles/sub-patterns in a way that
(i) 
pixels of the two tiles are overlaid if possible (no conflicts in color),
but not necessarily to the maximal extent,
(ii)
depending on further constraints, 
the uncovered space between the two sub-patterns can be minimized,
or the cover level (overlay level) aimed at can be minimized or maximized.

Fig.~\ref{MakeLoop}
shows the mini-loop embedded into an array of size $4 \times 4$
including border. 
Zero-pixels of the tiles are allowed to cover the border cells with the constant value zero by definition. 

\begin{figure}[H] 
\centering
\includegraphics[width=5cm]{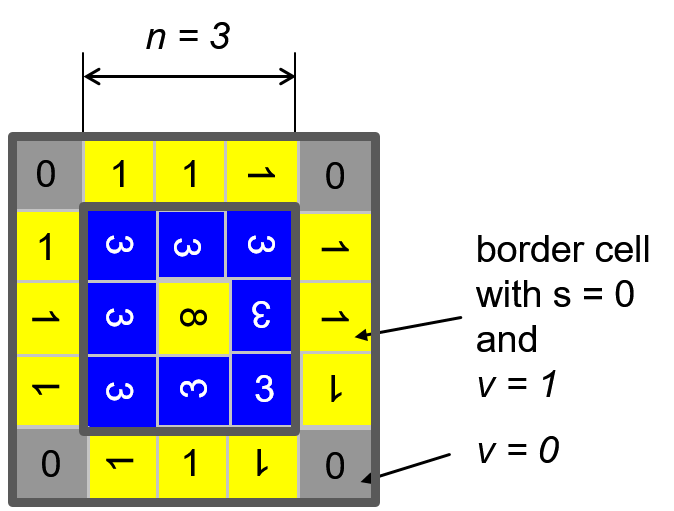}
\caption{
Mini-loop embedded in a 2D field of size $4 \times 4$.
The active area is $3 \times 3$.
}
\label{Loop3x3Embedded}
\end{figure}

\subsection{The derived templates}
\label{The derived templates}

\begin{figure}[H] 
\centering
\includegraphics[width=10cm]{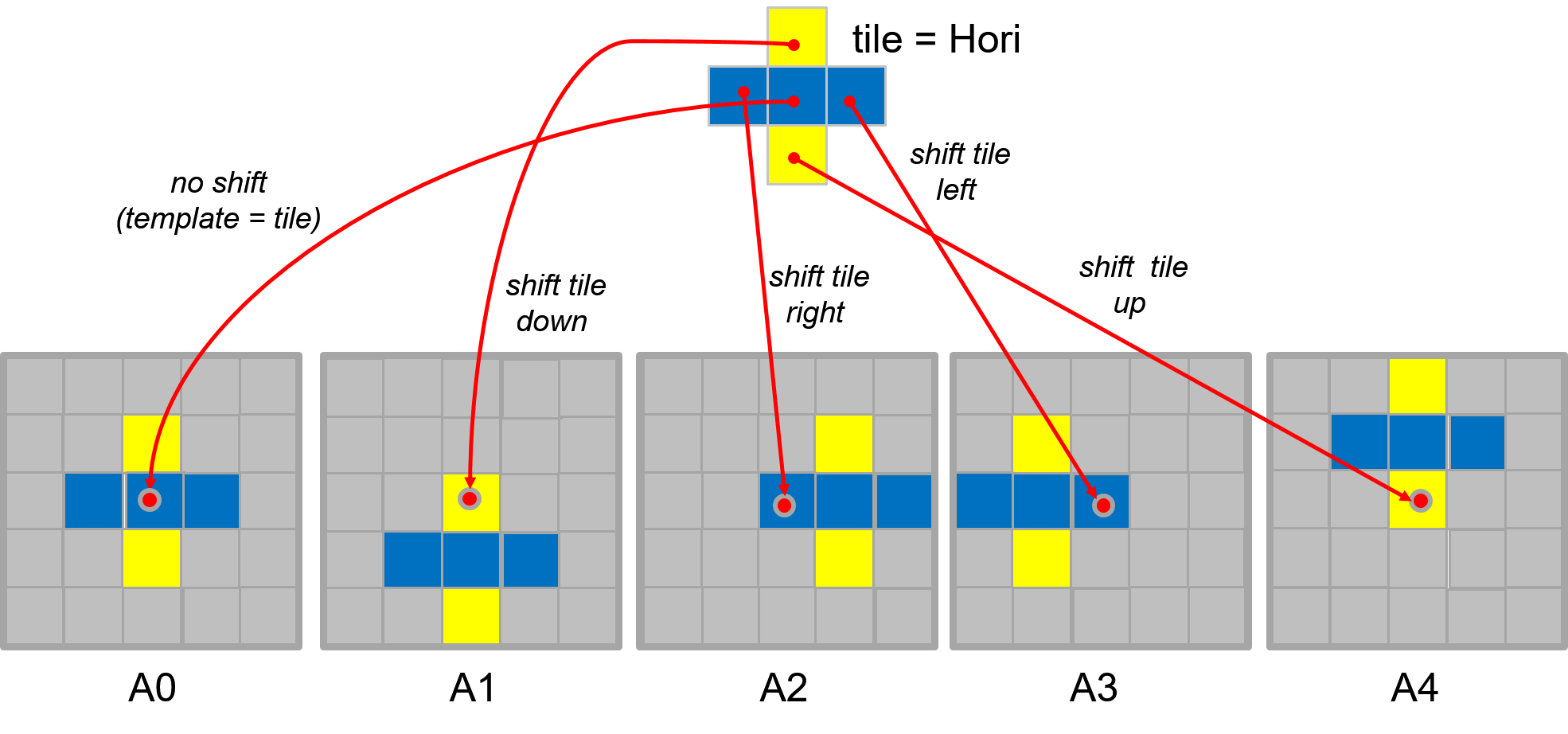}
\caption{
The five templates derived from the tile 'Hori'.
}
\label{DerivedTemplates}
\end{figure}

What is a \textit{template}?
A template is a local matching pattern.
It defines local conditions/constraints for valid patterns.
For each tile $T$ of the tile set,
$K$ templates $T_0, T_1, ... T_{K-1}$ are derived, where $K$ is the number of valid tile pixels with the value 0 or 1. 
The template $T_i$ is derived by shifting the tile in way that the valid pixel $i$ becomes the center of that template.
By convention we define $T_0=T$, i.e. the template $T_0$ is equal to the tile without shifting.
Note that we assume a center/anchor for each tile, and also for each template. 
Fig.~\ref{DerivedTemplates} illustrate this process for the tile 'Hori'.
Five templates A0, A1, A2, A3, A4 are derived by shifting, where A0 = Hori. 

\begin{figure}[H]
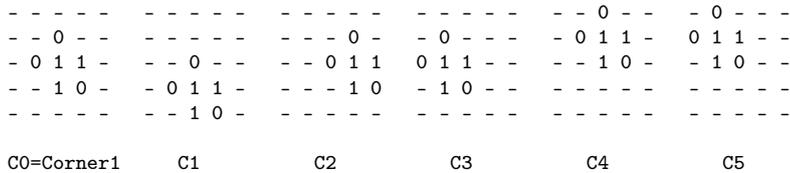
 
\footnotesize
\begin{verbatim}
- - - - -   - - - - -   - - - - -   - - - - -   - - 0 - -   - 0 - - -   
- - 0 - -   - - - - -   - - - 0 -   - 0 - - -   - 0 1 1 -   0 1 1 - -
- 0 1 1 -   - - 0 - -   - - 0 1 1   0 1 1 - -   - - 1 0 -   - 1 0 - -
- - 1 0 -   - 0 1 1 -   - - - 1 0   - 1 0 - -   - - - - -   - - - - -
- - - - -   - - 1 0 -   - - - - -   - - - - -   - - - - -   - - - - -

C0=Corner1     C1          C2          C3          C4          C5      
\end{verbatim}
\normalsize
\caption{
The six templates derived from Corner1.
The symbol '-' denotes sites which are don't cares where no valid pixel exists
and where no test is performed. 
}
\label{CornerTemplates}
\end{figure}

In then same way the templates for the other tiles are derived.
For instance, the templates derived from Corner1 are shown in 
Fig.~\ref{CornerTemplates}.
The number of all templates becomes $2 \times 5 + 4 \times 6 = 34$.

\section{The CA rule}
\label{The CA rule}

\subsection{Cell state}
\label{Cell state}

The used CA cell state at the site $(x,y)$ is

$q=(s, h_0,h_1)$ , 

where $s \in \{0,1\}$ is the main pattern state (color), 
and 
$h_0 \in \{0,1, ... , 8\}$ is a template hit with $s=0$,
and
$h_1 \in \{0,1, 2,3\}$ is a template hit with $s=1$
and $(x,y)$ are the coordinates (indexes) of the cell. 

The hits $h_0, h_1$ can be seen as temporary values,
they need not to be stored between micro-steps or generations, 
although their stored values can be useful for information/analytics 
or even needed in more complex rules.
It can be noticed that the hit value converges to the cover level
when a pattern stabilizes. 

\subsection{Asynchronous updating}
\label{Asynchronous updating}

\begin{figure}[H] 
\centering
\includegraphics[width=11cm]{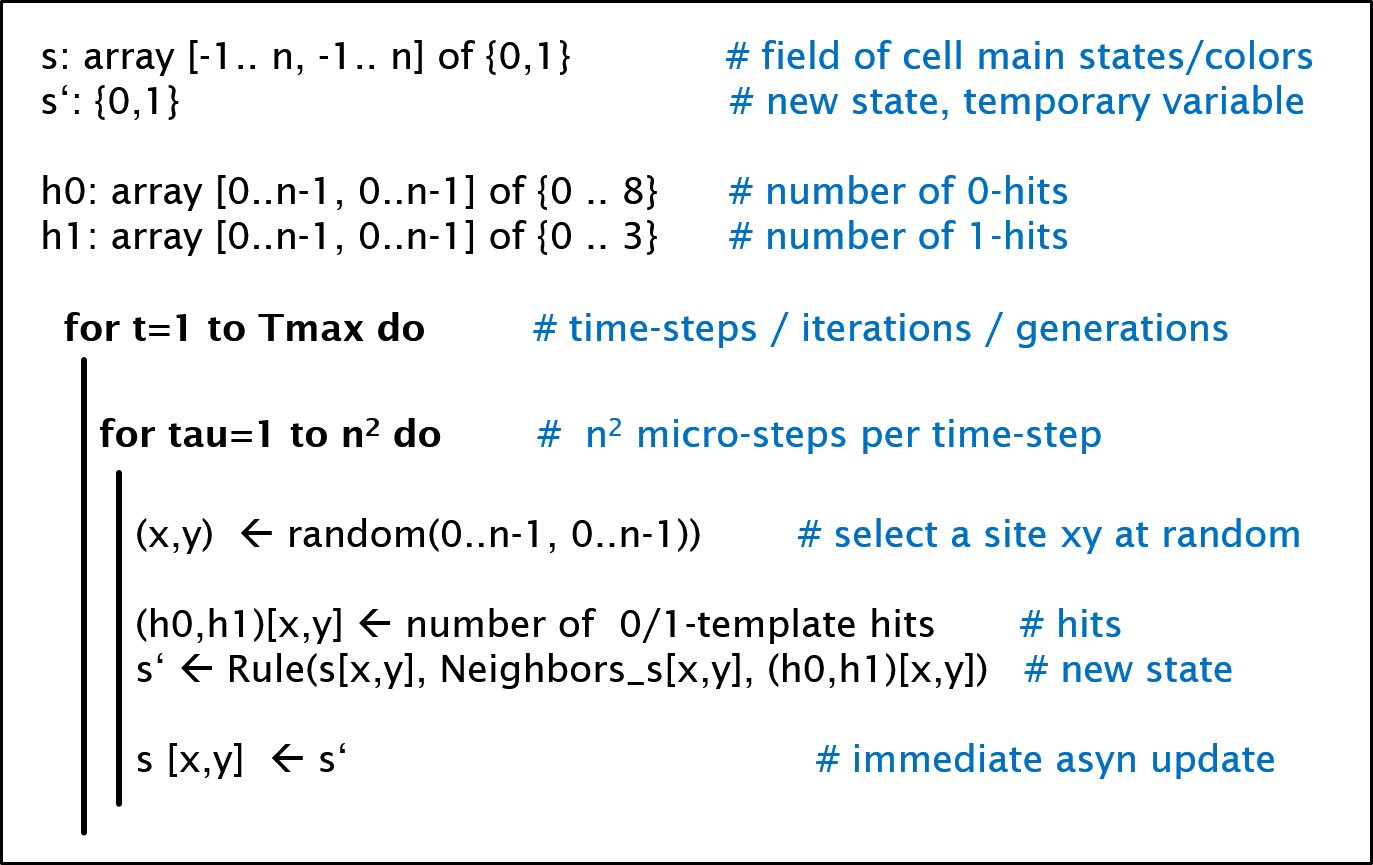}
\caption{
Algorithm for computing asynchronously $T_{max}$ generations.
For each generation $n^2$ cell operations (micro-steps) are executed.
}
\label{AsynUpdating}
\end{figure}

We want to use asynchronous execution/simulation of the CA rule
as shown in the Algorithm (Fig.~\ref{AsynUpdating}).
We use a simple scheme with the following cell operation
at each micro-step

\begin{enumerate}
	\item 
  A cell \texttt{(x,y)} is picked at random, 
  
  \item
  The new state value \texttt{s'} is computed by the CA rule
  using the parameters

  \texttt{s[x,y]~~~~~~~~~~~~~} (cell's main state/color)
  
  \texttt{Neighbors\_s[x,y]~~~} (the neighboring colors)
  
  \texttt{(h0,h1)[x,y]~~~~~~~} (the number of hits).
  
  \item
  The new state value \texttt{s'} is immediately copied to the state variable \texttt{s[x,y]}
  yielding the next state.
\end{enumerate}

For each generation (time-step $t \rightarrow t+1 $),
$n^2$ cell operations (micro-steps $\tau \rightarrow \tau +1$) are executed. 
For output it is useful to compute also the cover level for each generation.
This can be done by testing at each site all tiles and storing the 
number of pixels that overlay.

\subsection{Rule details}
\label{Rule details}

The rule is composed of two function evaluations (steps) in sequence.
In the first step a template test is performed. 
In the second step the new state is computed depending on the test result.
Note that the rule as a whole forms a function although it is computed stepwise. 

\begin{enumerate}
	\item TEST for template hits
	
	All templates are tested against the current pattern at site $(x,y)$ 
  where the center of each template is aligned to this site.
	The test is performed only on the outer neighborhood without considering the central state $s(x,y)$.
	So the current value of the cell is not taken into account, only the real neighbors. 
		
	$(h_0, h_1, k) \leftarrow \textit{ComputeHitsForAllTemplates}~ T_i \in \{T_0, T_1, ... T_{K-1}\}$
	
	where
	
	$h_0$ = number of zero-hits (hits with template center value = 0)
	
  $h_1$ = number of one-hits (hits with template center value = 1)
	
  $k$ = index of a matching template $T_k$ 
	
	$K= 34 = 4 \times 2 + 6 \times 4$, the number of templates (\#valid pixels in all tiles).

  \item NEW STATE depending on test result
  
  The default new state is $s'=s$ (keep old state).
  If there is a hit, the state is adjusted to the center of a matching template,
  otherwise noise is injected. 
  
  
    \begin{itemize}
    
      \item ADJUST IF HITS
      
      $s' = center(T_k) ~~~ \text{if~ } h= h_0+h_1 >0$
      
      where 

      $center(T_k)$ is the value of the center pixel  of a matching template $T_k$.

      If there is only one hit, the cell's state $s$
       is adjusted to the value of the template's center value. 
      This means that the center color is corrected if all other pixels of the template are correct.
      In the case when the whole template (including the center) correctly matches,
      the adjustment is not necessary, it is redundant.
      If there are several hits, one of the hitting templates is selected at random for adjustment. 
      If there is no hit, then execute the next step.
      
      \item NOISE INJECTION IF NO HITS
          
      $s' =
      \begin{cases}
       R_p    & \text{if~ } P \wedge (h=0)\wedge (s=1) \wedge \textit{TRUE}(\pi_1) \\     
       R_q    & \text{if~ } Q \wedge (h=0)\wedge (s=0) \wedge \textit{TRUE}(\pi_0) \\
      \end{cases}$
      
      where
      
      $P=(h_1 \neq 3)$      ~~~~-- the 'path condition' for a closed loop
          
      $Q= (h_0=0)$          ~~~~-- this conditions forbids uncovered cells
      
      $R_P$ = Random(0/1),      
      $R_Q$ = Random(0/1)
      
      \textit{TRUE}($\pi$) $\Leftrightarrow$ \textit{TRUE} with probability $\pi$ else \textit{FALSE}
            
    \end{itemize}
    	
\end{enumerate}

\section{Evolved patterns}
\label{Evolved patterns}

The CA rule was used with the parameters $\pi_1=0.5, \pi_0=0$.
The CA  rule evolves loop patterns with
(a) no-touching loops, or
(b) patterns with some loops that are touching, or
(c) no loops at all, all cells are in state 0 (all-zero), and no cell is covered.

The case (c) is due to $\pi_0=0$. It turned out that with this parameter setting the all-zero state appears 
very rarely, but the evolution is faster compared to $\pi_0>0$.
Setting $\pi_0>0$ would exclude patterns with uncovered cells,
like 
(Fig.~\ref{Pattern4x4}a,
Fig.~\ref{OldTileNewTileLoop}, 
Fig.~\ref{TouchingOnlyLoops5x5}a1 -- a5, b3 -- b5).
Then such interesting loop patterns with uncovered cells would be destroyed  during the evolution
and could appear only as transients. 
Therefore we used  $\pi_0=0$.

\subsection{Patterns of size $4\times 4$}
\label{Patterns of size 4}

\begin{figure}[H] 
\centering
\includegraphics[width=0.8\textwidth]{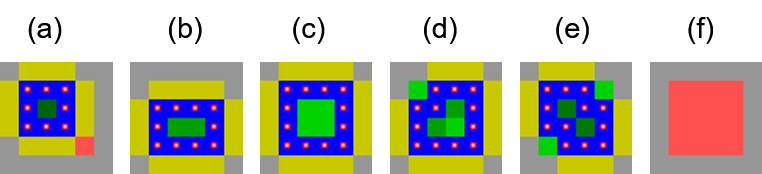}  
\caption{
The five possible $4\times 4$ loop patterns.
(a) Is the simplest loop, with one uncovered cell (in red),
(a -- d) are non-touching loops, 
(e) is a self-touching loop, touching internally, 
(f) is the all-zero case, rarely appearing.}
\label{Pattern4x4} 
\end{figure}

If the size of the field is $n^2=3\times3$ then there can only be one loop,
the simple mini-loop (Fig.~\ref{Loop3x3Embedded}).
For $n=4$ there are five possible loop patterns (Fig.~\ref{Pattern4x4}).
The first one (a) is the mini-loop already known.
There are four new loops (b -- e).
Loops (a -- c) have 4 corners, where the corners point to the exterior area outside
the loop. Such corners are attributed as ``convex''.
Loop (d) has 6 corners, 5 convex  and one ``concave''
(pointing to the interior enclosed area). 
So loops (a -- c) can be classified as \textit{convex}, and
loops (d, e) as \textit{concave}, because they contains at least one concave corner.
Loop (d) has 6 convex and 2 concave corners,
and it is the only self-touching loop with one touch point.

\begin{figure}[H] 
\centering
\includegraphics[width=0.8\textwidth]{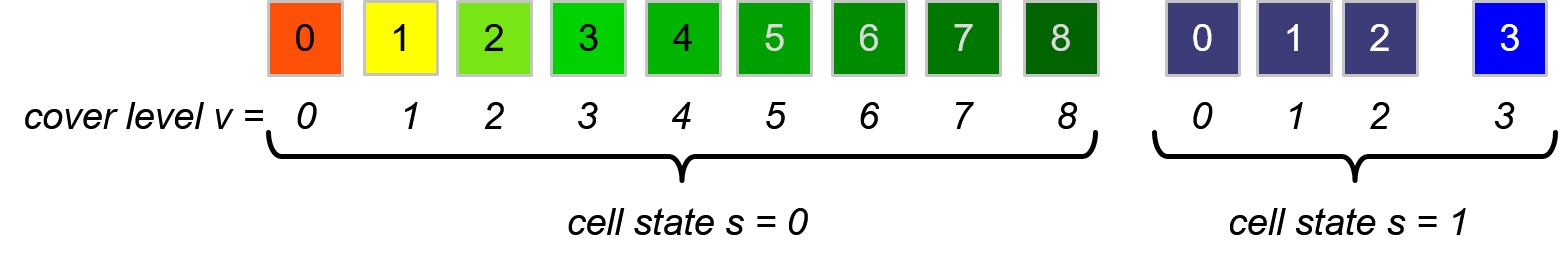}  
\caption{
The coloring scheme used for most of the shown patterns.}
\label{Coloring} 
\end{figure}

\textbf{Coloring.}
Most of the shown patterns use the coloring scheme as depicted in 
Fig.~\ref{Coloring}.

\subsection{Patterns of size $5\times5$}
\label{Patterns of size 5}

In the former publication 
\cite{2024-Hoffmann-Bialecki-Loop-AdvancesInCA}  
all the non-touching loops were already found by a similar CA rule
with a different corner tile.
In the new corner tile, the pixel at the outer diagonal of the corner 
point was omitted (Sect. \ref{SectSetOfTiles}).
All possible loop patterns of size $5 \times 5$ are:

\textit{\textbf{Citation (taken from \cite{2024-Hoffmann-Bialecki-Loop-AdvancesInCA}) begin.}}

\begin{figure}[H] 
\centering
\includegraphics[width=\textwidth]{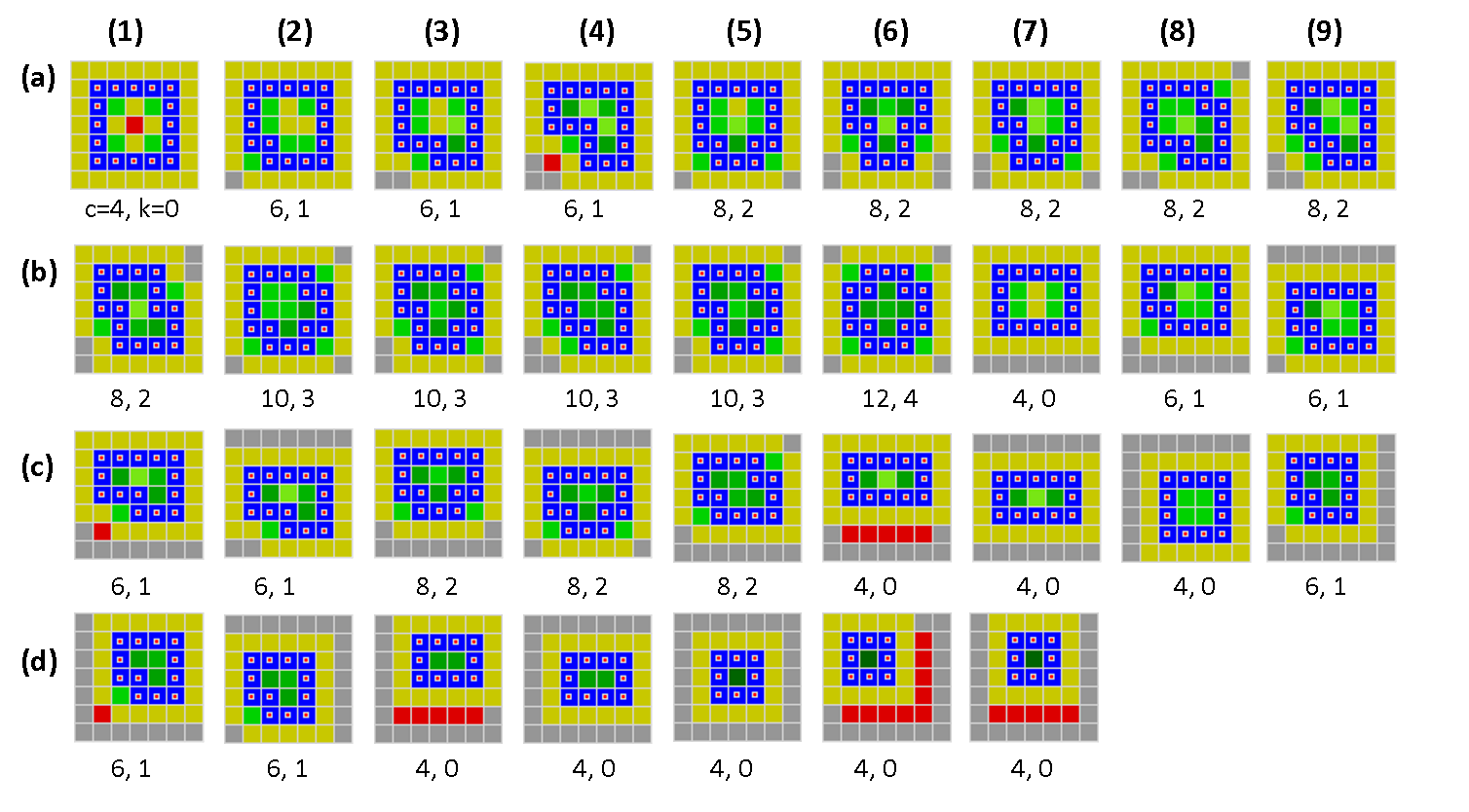}  
\caption{
All the detected loop patterns of size $5\times 5$. 
Uncovered cells, which are not on the boundary, are marked in red. The number of corners $c$ and 
the number of inner/convex corners $k$ are given.
The following loops are equivalent under shift:  
$(b8, b9), (c1, c2), (c3, c4), (c6, c7), (c9, d1, d2), (d3, d4), (d5, d6, d7)$.
The patterns are ordered with respect to  
	the size of the rectangle  that encloses a loop.} 
	\label{ALL5x5} 
\end{figure}

For a field of size $5\times5$ we found  33 possible loop patterns (Fig.~\ref{ALL5x5}),
where symmetric patterns under rotation and mirroring are not shown/counted. 
Some of the loops show the same structure but are shifted which results in a different 
covering configuration, often with uncovered cells (marked in red if not on the boundary). 
If patterns with the same loop structure (equivalents under shift) are counted only once
then we obtain only 24 different loops.
%
%
\textit{\textbf{Citation end.}}

\begin{figure}[H] 
\centering
\includegraphics[width=3.5cm]{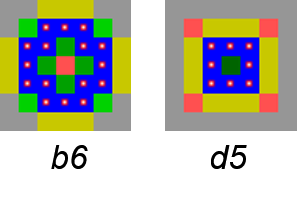}
\caption{
The cover levels are lower at certain sites if the new corner tile is used instead of the old one.
Two loops are shown generated with the new corner tile. 
(They can be compared with the similar ones depicted in Fig.~\ref{ALL5x5}.)
}
\label{OldTileNewTileLoop}
\end{figure}

It should be noted that the cover levels differ when the new resp. the old tile
is used (compare Fig.~\ref{OldTileNewTileLoop} with Fig.~\ref{ALL5x5}).

\textbf{Touching loops of size $5\times 5$.}   
There are 8 possible touching loops, counting symmetric and translational ones only once. 
Loop (a1, a2) is already known as $4\times 4$ loop.
Loops (a3, a4) are equivalent under shift. 
Loop (b5) has two touching points, whereas the others have only one,
and it has the maximum of two self-touching points
and 4 convex corners.

\begin{figure}[H] 
\centering
\includegraphics[width=8cm]{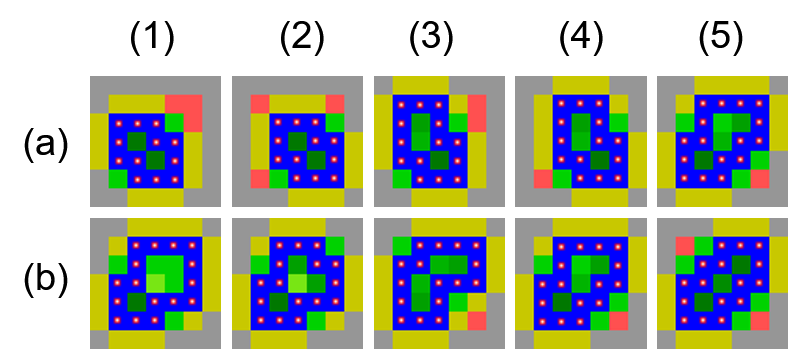}
\caption{
The touching loops evolved by the CA rule.
The patterns $(a1,a2),(a3,a4)$ are equivalent under shift. 
There are 8 different loops, counting equivalents under shift and symmetry only once. 
}
\label{TouchingOnlyLoops5x5}
\end{figure}

\subsection{Patterns of size $6\times6$}
\label{Patterns of size 6}

\begin{figure}[H] 
\centering
\includegraphics[width=\textwidth]{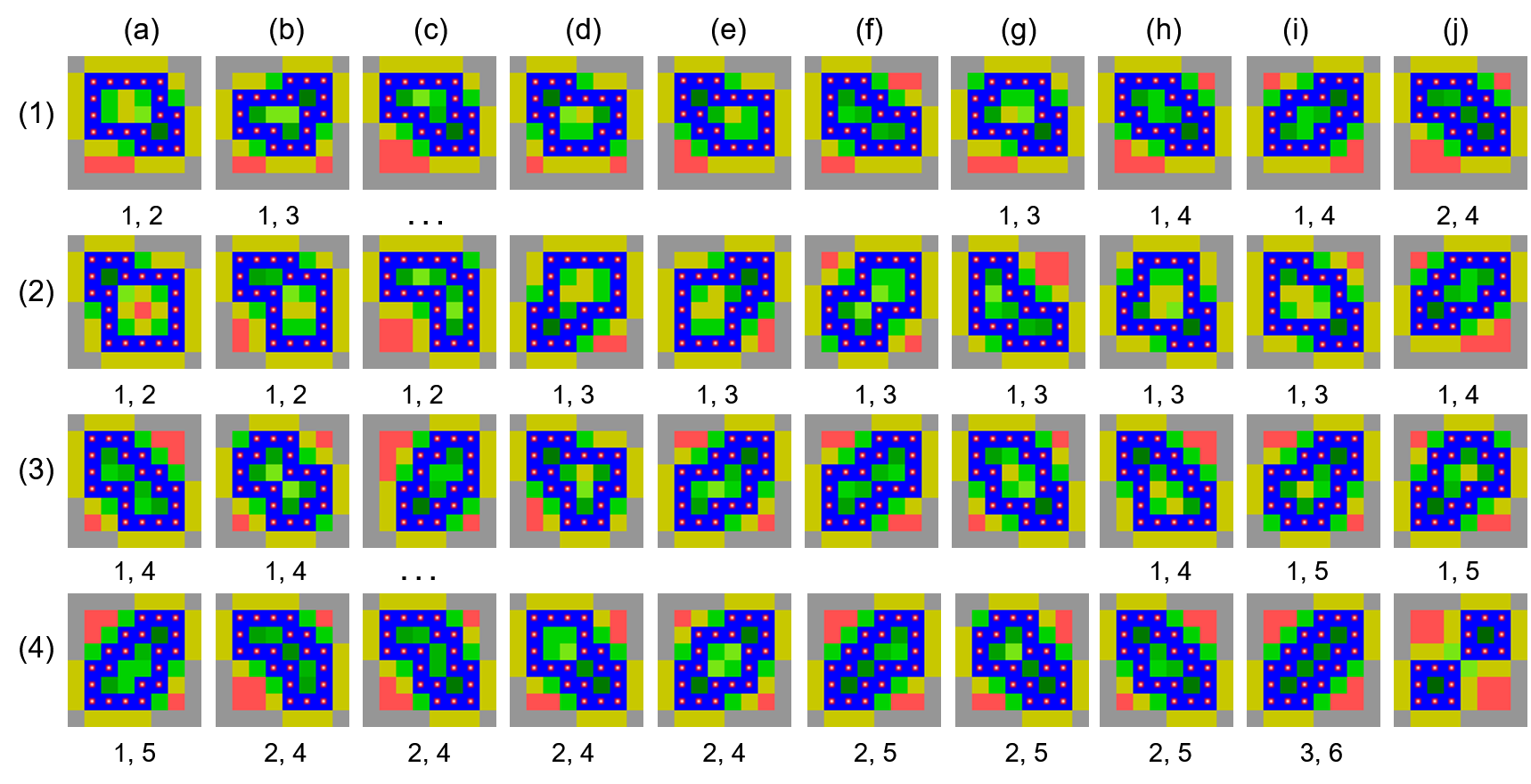}
\caption{
A selection of  $6\times6$ touching patterns.
}
\label{Touching6x6}
\end{figure}

A selection of touching loops is shown in Fig.~\ref{Touching6x6}.
The first index under a pattern gives the number of touching points,
and the second index gives the number of concave corners. 
The first row (1) shows loops fitting into a $5\times6$ area.
Most frequently pattern with indexes (1,3) and (1,4) were evolved. 
Patterns 4i and 4j appear very rarely. 
Pattern 4i is a fully self-touching loop in which the number of
touching points is maximal  and half of the number of concave corners.
Pattern 4j is the only pattern with two loops with an external touching point. 

\subsection{Patterns of size $7\times7$}
\label{Patterns of size 7}
\begin{figure}[H] 
\centering
\includegraphics[width=\textwidth]{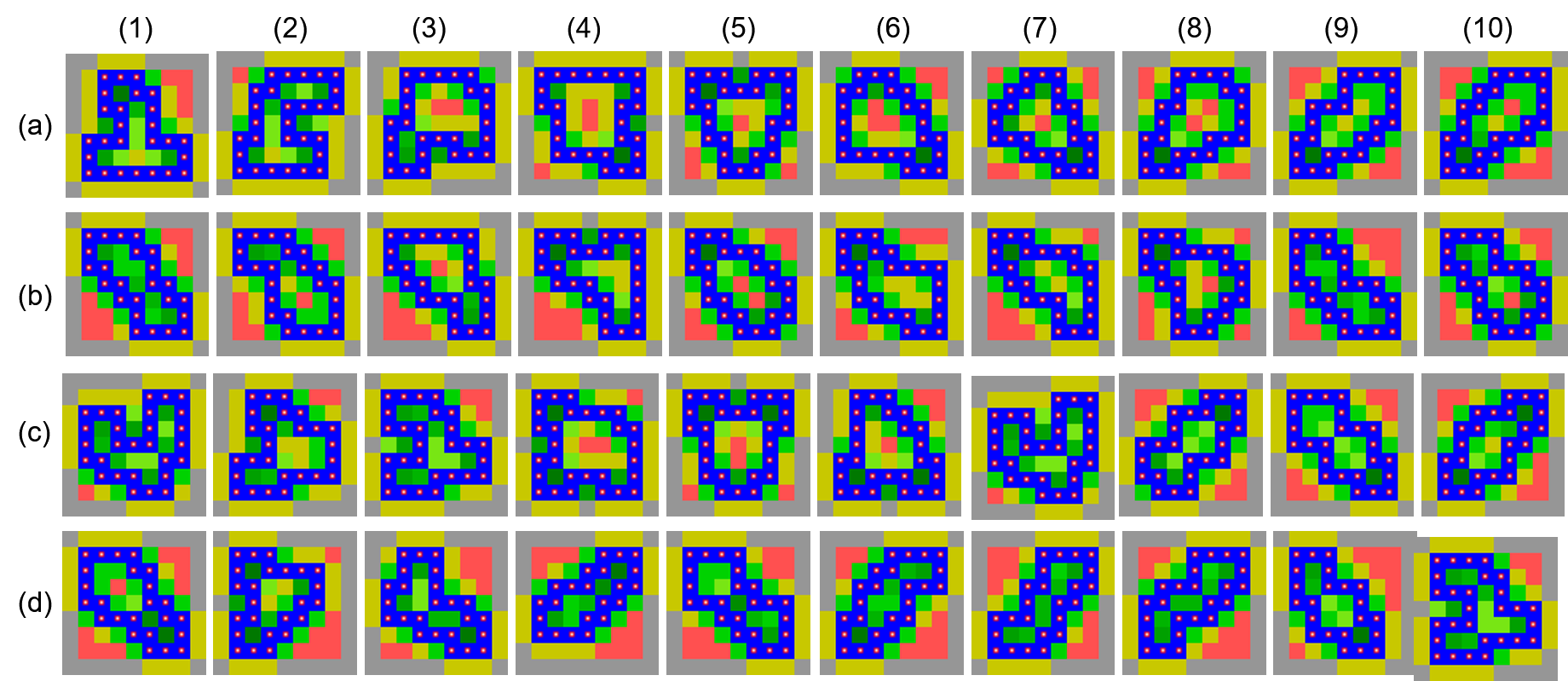}
\caption{
A selection of  $7\times7$ loops,
(a, b) with one touching point and
(c, d) with two touching points.  
}
\label{Patterns7x7}
\end{figure}

\begin{figure}[H] 
\centering
\includegraphics[width=\textwidth]{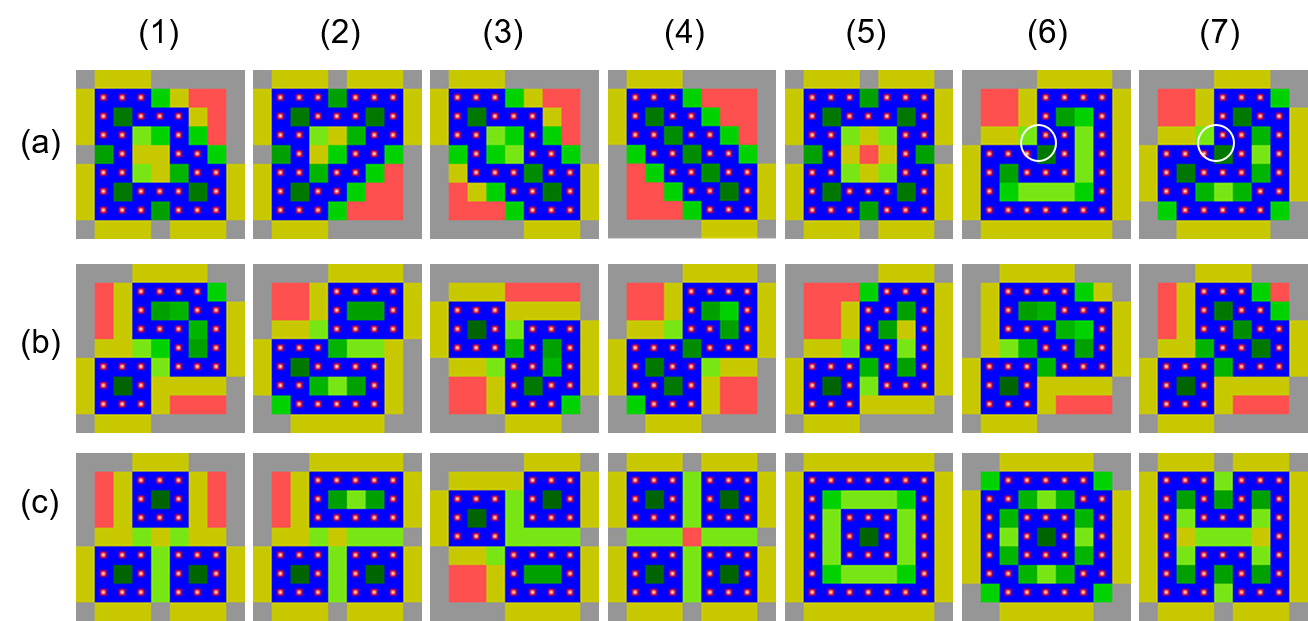}
\caption{
Special $7\times7$ patterns.
(a1, a2, a3)/(a4, a5) loops with 3/4 self-touching points. 
Loops a5 and a7  each have cells that touch each other from the
outer side of the loop's path. 
}
\label{SpecialPatterns7x7}
\end{figure}

\begin{figure}[H] 
\centering
\includegraphics[width=0.8\textwidth]{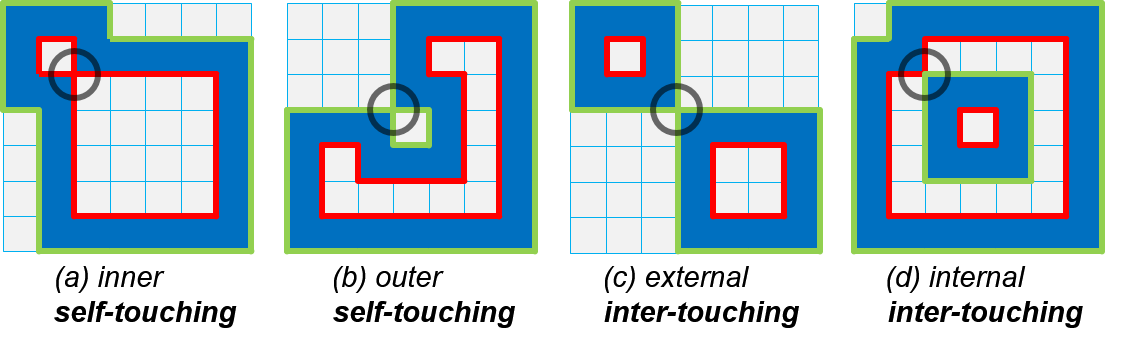}
\caption{
Different types of touching-points.
A loop has an inside (red) and an outside (green).
There are four different types of touching-points depending on 
whether cells are touching with the inside or the outside
of the path.
}
\label{SelfTouching}
\end{figure}

A selection of evolved $7\times7$ loop pattern is shown in Fig.~\ref{Patterns7x7}.
The patterns are sorted by the number of self-touching points. 
Loops with 3 or 4 self-touching points appear rarely.
Special $7\times7$ patterns are shown in Fig.~\ref{SpecialPatterns7x7}.
The loop c7 is a non-touching space-filling curve, here just presented as a general information. 
Loops with 3 or 4 self-touching points appear rarely.
A few of the shown patterns were constructed (but are stable and can evolve in principal),
namely a4, a6, a7, c5, c6, c7.
A loop encloses another in c5 and c6, where there are four touching points between the loops in c6.
Loops a6 and a7 have an entry-point (bottleneck) from outside the loop to a cave, 
there are cells which touch each other with the outside of the loop's path.
Such outer-touching points can appear for $n\geq7$.
The loops c4 -- c7 are space-filling, the number of path cells is maximal (32).
We can distinguish four different point types:

\begin{itemize}
	\item \textit{self-touching point}
    
    \begin{itemize}      
      \item \textit{inner}      
      
      Two cells are touching with the \textit{inner} side of the loop path.      
      Such a point is called  ``\textit{inner self-touching point}''.
      
      \item \textit{outer}  
      
      Two cells are touching with the \textit{outer} side of the loop path.        
      Such a point is called  ``\textit{outer self-touching point}''.
      Such points lie in concave loop segments and are bottlenecks within caves or bottles,      
      as exemplified in Fig.~\ref{SpecialPatterns7x7}(a6),
			Fig.~\ref{SelfTouching}(c), 
      Fig.~\ref{Pattern8x8}(e10).   
           
    \end{itemize}   
  
  \item \textit{inter-touching point}  
      
      \begin{itemize}
        
				\item \textit{external}
				
        Two cells of different loops A and B are touching. 
        If both outer sides of the loops' paths are touching, then
        the point is called ``\textit{external inter-touching point}''.        
        This is the more frequent case when loop A is situated alongside loop B.
        
        \item \textit{internal}				
				
        Two cells of different loops A and B are touching
        where A encloses B.
        Then the inner side of path A touches the outer side of path B. 
        Such point is     
        called ``\textit{internal inter-touching point}'', or
        ``\textit{enclosing inter-touching point}''.        
        
      \end{itemize}
\end{itemize}

\begin{figure}[H] 
\centering
\includegraphics[width=\textwidth]{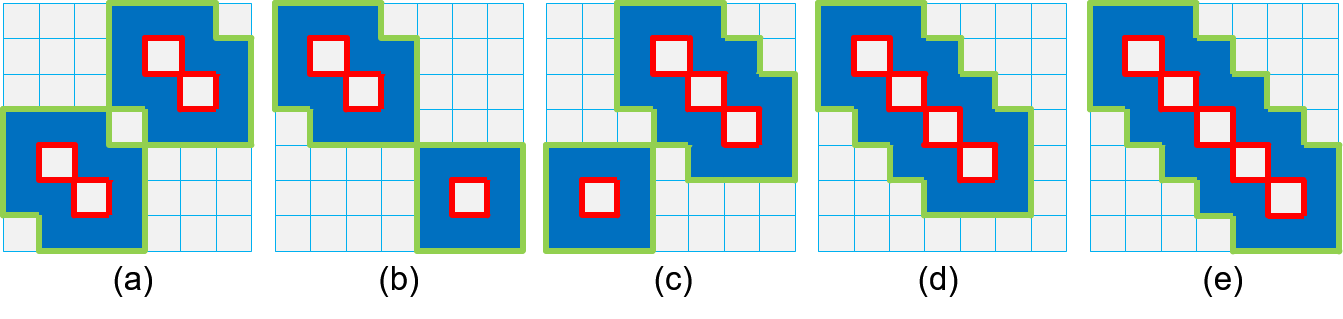}
\caption{
Special $7\times7$ patterns with ``worms''.
The largest possible worm for $n=7$ is  a 4-worm (e).
}
\label{Worm7x7}
\end{figure}

\textbf{Worms.}
We define a ``\textit{Worm}'' as a loop that is strongly contracted with a maximal number 
of inner self-touching points, and it forms a diagonal of zeroes enclosed by ones.
We do not consider the simple square $3\times3$ loop (mini-loop) as a worm.
A \textit{k-worm} is a worm that has $k \geq 1$ inner self-touching points. 
The length of the zero-diagonal is $k+1$.
A $k$-worm fits into a square grid of size $n^2 = (k+3)^2$,
$k=n-3$.
Some $7\times7$ patterns with worms are shown in 
Fig.~\ref{Worm7x7}.

\subsection{Patterns of size $8\times8$}
\label{Patterns of size 8}

\begin{figure}[H] 
\centering
\includegraphics[width=\textwidth]{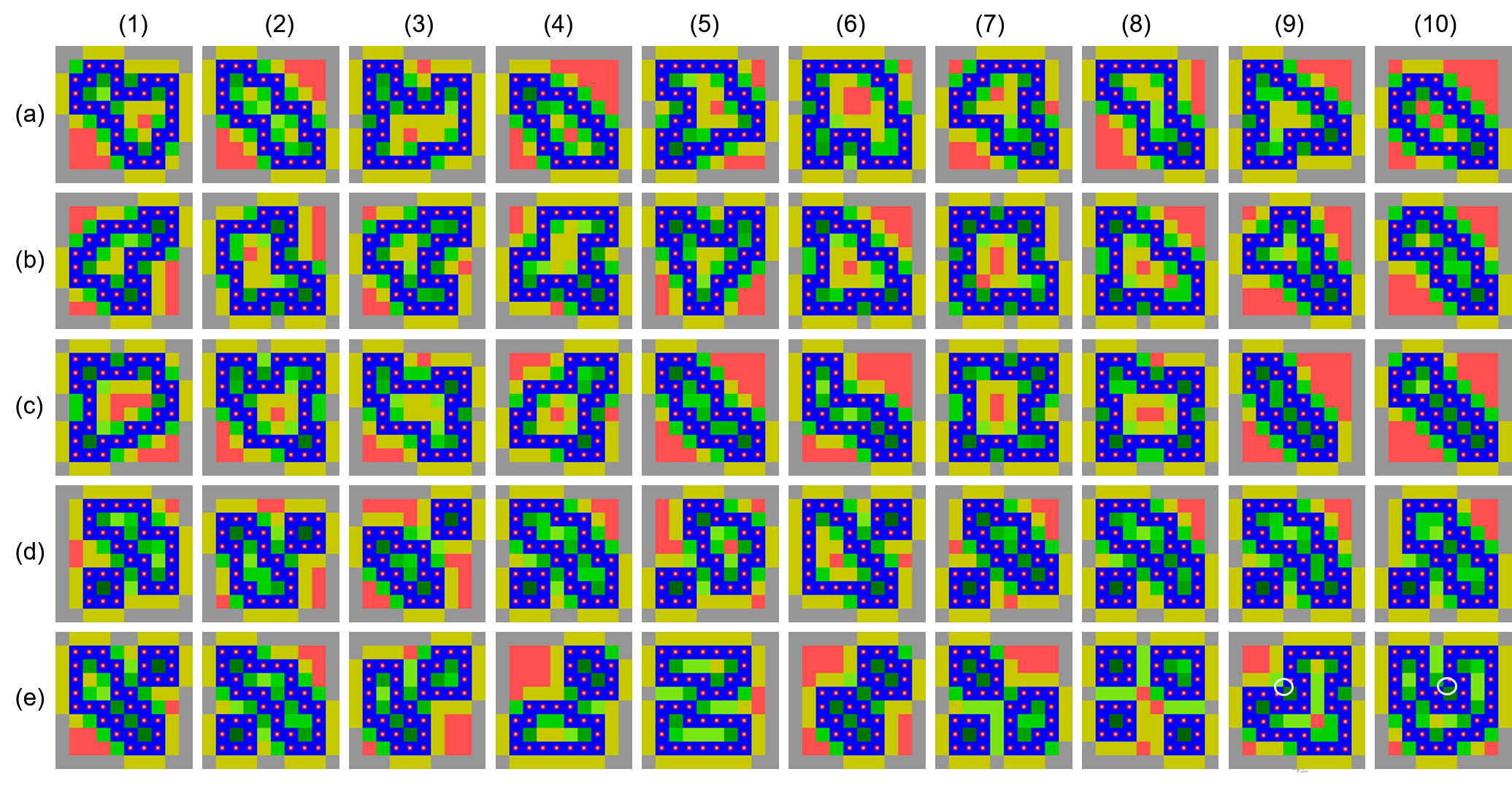}
\caption{
A selection of evolved $8\times8$ patterns, showing loops with
(a) one self-touching point,
(b) two, 
(c1 -- c8)  three, and
(c9, c10) four. 
Patterns (d1 -- e7) each show two external inter-touching loops.
Loops e9 and e10 each contain an outer self-touching point.  
}
\label{Pattern8x8}
\end{figure}

A selection of evolved $8\times8$ loop pattern is shown in Fig.~\ref{Pattern8x8}.
Row (a) shows 10 loops with \textit{one} self-touching point,
and row (b) shows 10 loops with \textit{two} self-touching points. 
Loops (c1 -- c8) have \textit{three} self-touching points,
and loops (c9, c10) have \textit{four}. 
Patterns (d1 -- e7) each show \textit{two inter-touching} loops,
two loops that touch externally.
Loops e9 and e10 each show a loop with an \textit{outer self-touching} point.  

\begin{figure}[H] 
\centering
\includegraphics[width=0.6\textwidth]{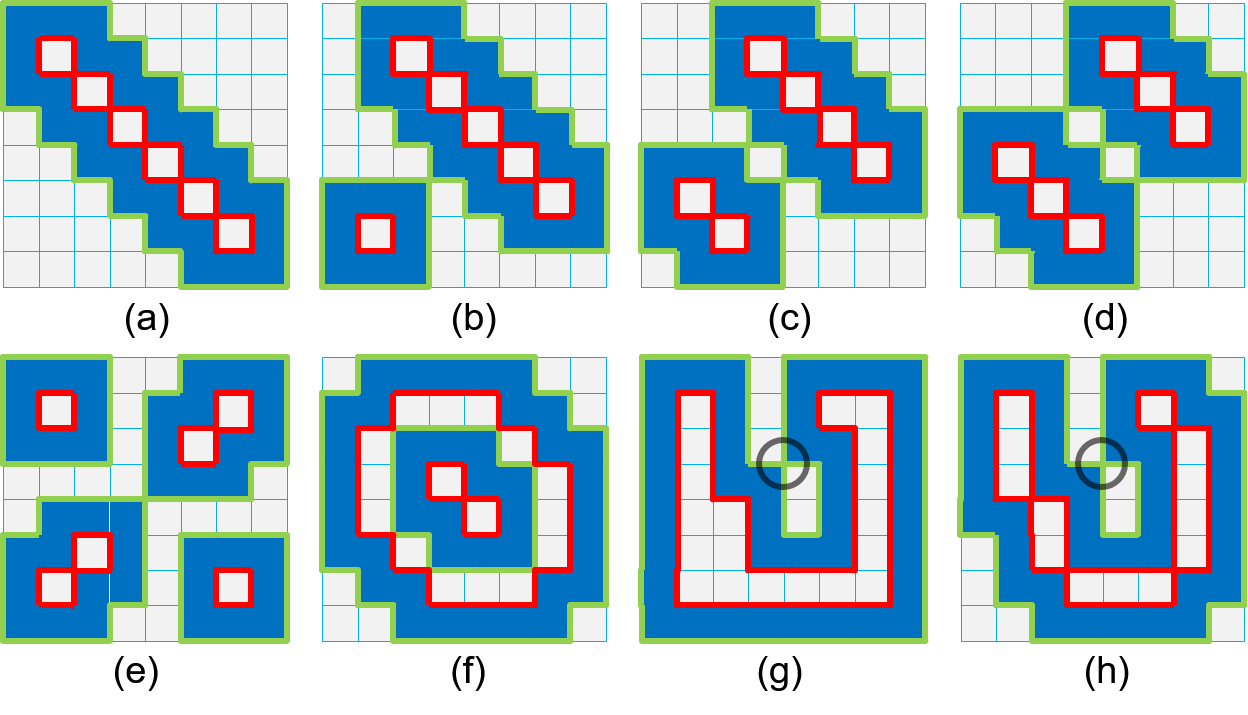}
\caption{
Special $8\times8$ patterns. 
(a -- f) Patterns with worms. 
Loops (g, h)  contain an outer self-touching point.
}
\label{Special8x8}
\end{figure}

The patterns (Fig.~\ref{Special8x8})
are constructed by hand although they can be evolved by the CA rule. 
The patterns (a -- f) contain worms, and (g, h) 
show a self-touching point, a bottleneck within a cave
 (a concave loop segment).

\subsection{Patterns of size $16\times16$}
\label{Patterns of size 16}

\begin{figure}[H] 
\centering
\includegraphics[width=0.6\textwidth]{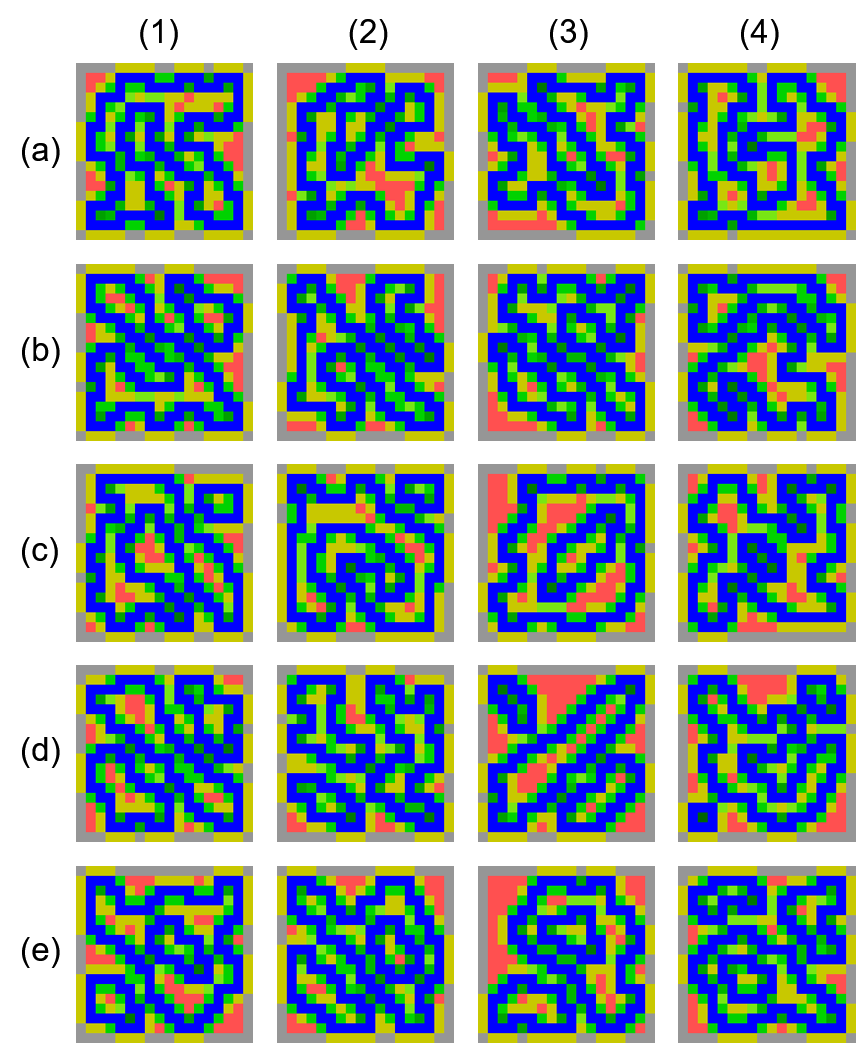}
\caption{
$16\times16$ evolved patterns. 
(a) Patterns with one loop,
(b, c) with two,
(d, e1 -- e3) with three, 
(e4) with four loops.
}
\label{Loops16x16}
\end{figure}

The CA rule can evolve very many different, stable touching loop patterns,
a selection is shown in 
Fig.~\ref{Loops16x16}.
The rate of evolved patterns with 
 one loop, with four (or more) was much lower than 
that for patterns with two or three loops. 
Special regular patterns (like the ones shown in 
Fig.~\ref{Worm7x7}
or
Fig.~\ref{Special8x8})
can appear but the probability is low because the amount of all valid loops patterns is
very high. 

\begin{figure}[H] 
\centering
\includegraphics[width=\textwidth]{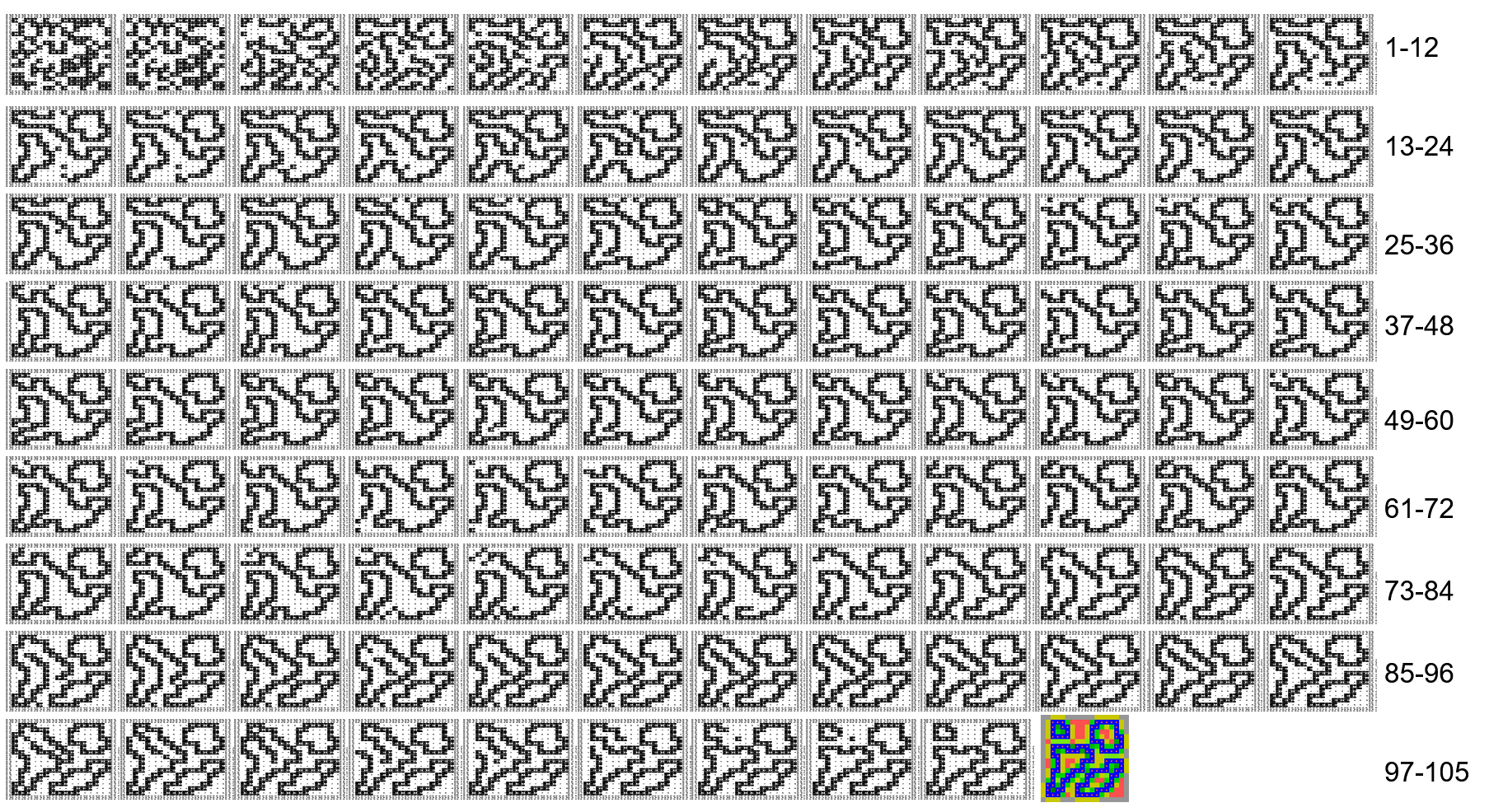}
\caption{
An example for a time-evolution of a stable loop pattern. 
}
\label{TimeEvolution16x16}
\end{figure}

The evolution of a stable loop pattern is relatively fast,
as we can see for an example
(Fig.~\ref{TimeEvolution16x16}).
After 10 -- 20 generations typical partial loop structures appear,
but it takes more time to change all structures into closed loops.  

\subsection{Time-complexity and density}
\label{Time-complexity and density}

\begin{figure}[H] 
\centering
\includegraphics[width=0.7\textwidth]{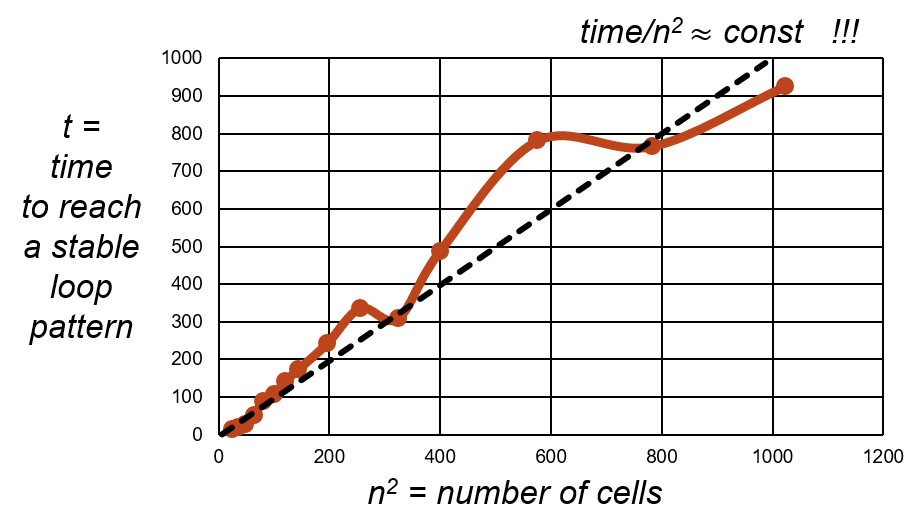}
\caption{
In the simulation, the time-complexity is approximately $O(n^2)$.
}
\label{TimeComplexity}
\end{figure}

\textbf{Time-complexity.}
The time-complexity was estimated by executing multiple runs
and averaging. 
The number of performed runs (each yielding a stable pattern) was
100 for $n=3, 4,  \ldots ,12$, and 
50  for $n=14, 16, 18, 20$, and
20  for $n= 24, 28$, and
10  for $n=32$.
Surprisingly, the time (time-steps \textit{t} needed for a stable loop pattern)
grows only approximately linear with the number of cells ($t=O(n^2)$),
as we may conclude from the curve's trend
(Fig.~\ref{TimeComplexity}).
This means that each cell consumes a constant number of time-steps.
(Here we got a special result where the constant is nearly 1.)
In the parallel case, when a fraction $\alpha$ of the cells are updated in parallel,
the parallel time could approximate a constant. To confirm this proposition more experiments and/or
theoretical investigations are needed.
From the simulation snapshots 
(Fig.~\ref{TimeEvolution16x16})
we can see that small loops appear relatively fast
(and probably survive),
and the largest loop needs more time (trials and error corrections) to evolve.
One explanation for the low time-complexity could be that a large loop can evolve faster in an  area
that was restricted by smaller, already formed loops. 

\textbf{Density.}
We define the density $\delta$ as the sum of all one-cells (path cells) divided by
$(n+1)^2$. We use $(n+1)^2$ instead of $n^2$ because then the maximal possible density 
of valid loop patterns
becomes $0.5$ when half of the cells is one and the other half is zero. 
This is the case for space filling curves. 
So we include half of the border cells into the number of all cells being considered,
the denominator of the formula

$\delta = {1 \over (n+1)^2} \sum_{\forall ij} (s_{ij}=1)$.

The maximal reachable density (upper bound) depends on the loop structure(s) and is
(where 'maximal loop' means a loop pattern with a maximal number of ones)

$\delta = 8/16=0.5$ for $n=3$, the smallest $3\times 3$ square mini-loop (Fig.~\ref{Loop3x3Embedded}),

$\delta = 10/25=0.4$ for $n=4$, maximal $4\times 4$  loops (Fig.~\ref{Pattern4x4}(c, d, e)),

$\delta = 16/36=0.44$ for $n=5$, maximal $5\times 5$  loops  
(Fig.~\ref{ALL5x5}(a, b1 -- b6), Fig.~\ref{TouchingOnlyLoops5x5}(a5, b)),

$\delta = 20/49=0.41$ for $n=6$, maximal $6\times 6$  loops  
(Fig.~\ref{Touching6x6}(subset of loops that have contact to all the 4 boundaries)),

$\delta = 32/64=0.5$ for $n=7$, maximal $7\times 7$  loops  
(Fig.~\ref{SpecialPatterns7x7}(c5, c6, c7)),

$\delta = 40/81=0.49$ for $n=8$, maximal $8\times 8$  loops  
(Fig.~\ref{Special8x8}(f, g, h)).

\begin{figure}[H] 
\centering
\includegraphics[width=0.8\textwidth]{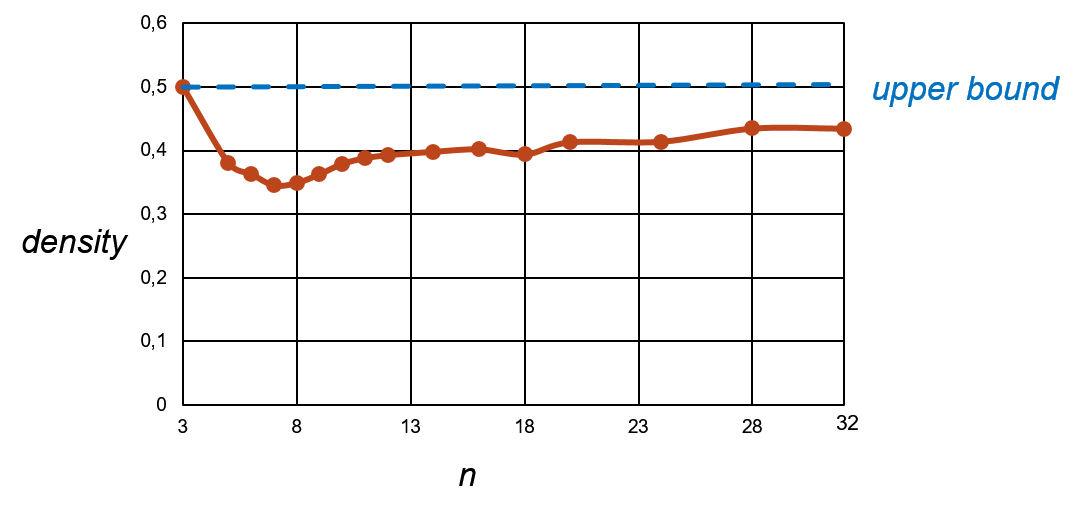}
\caption{
The density (averaged over the runs/evolved patterns) depending on the size of the grid. 
}
\label{Density}
\end{figure}

The density of the evolved patterns on average is shown in 
Fig.~\ref{Density}.
For $n=32$ the average density is $\delta = 0.435~ (87\%)$
The upper bound is $0.5$ which can only be reached for certain $n$ (e.g. 3, 7)
and patterns with a maximum of 1-cells. 

\subsection{Larger loops}
\label{Larger loops}

\begin{figure}[H] 
\centering
\includegraphics[width=\textwidth]{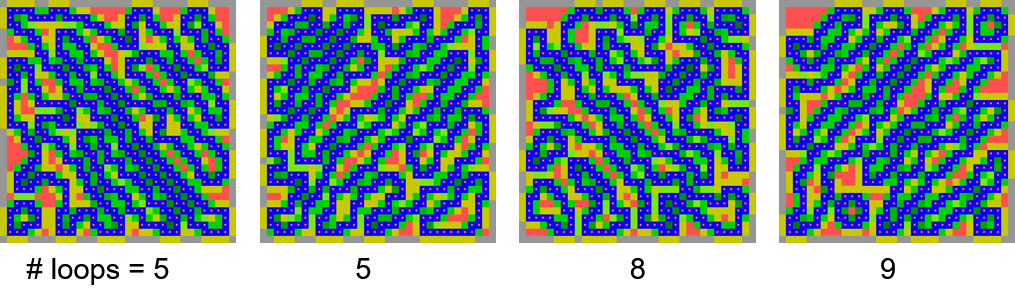}
\caption{
Some loop patterns of size $32\times32$ with a different number of loops.
}
\label{Loop32x32}
\end{figure}

\begin{figure}[H] 
\centering
\includegraphics[width=\textwidth]{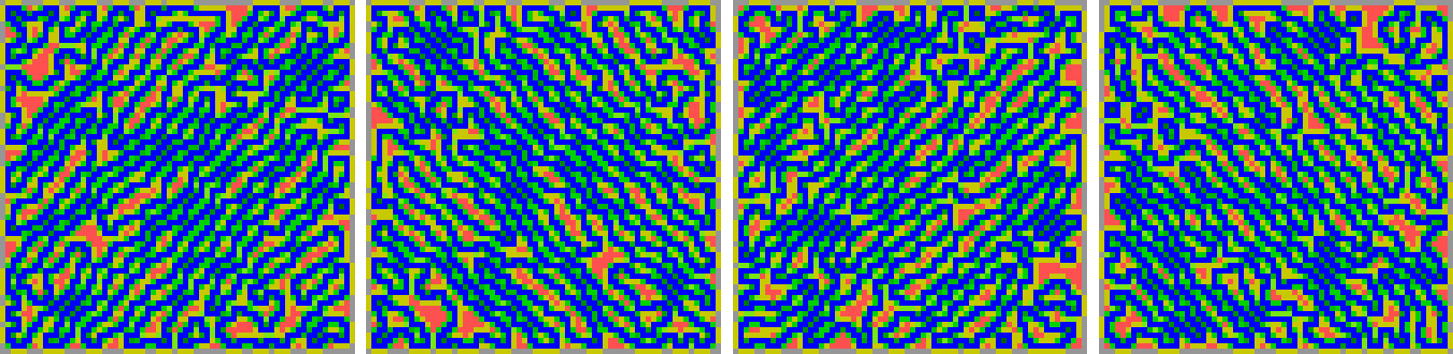}
\caption{
Examples for loop patterns of size  $64\times64$.
}
\label{Loop64x64}
\end{figure}

The CA rule is able to evolve also larger patterns with $n > 16$
with a reasonable computational effort
because of the low time complexity (Sect. \ref{Time-complexity and density}).
For illustration
we show only a few evolved patterns of size 
$32\times 32$ (Fig.~\ref{Loop32x32})
and
$64\times 64$ (Fig.~\ref{Loop64x64}).

\section{Conclusion and future work}
\label{Conclusion and future work}

A CA rule was designed that can evolve stable touching loop patterns.
First a set of 6 tiles 
(Hori, Verti, Corner1 -- Corner4) was defined. 
Each tile defines a  $3\times 3$
local pattern consisting of 0/1-pixels.
34 templates were  derived from the tiles by shifting them
in a way that each pixel becomes the center of a template.
The templates were inserted into a probabilistic CA rule with asynchronous updating.
All templates are tested for matching in the neighborhood of each CA cell.
The rule tries to establish a template at each site
by correcting the cell's value to the center value of a hitting template
or noise injection otherwise. 
In addition the path conditions needs to be fulfilled,
saying that each path cell shall have an overlap level of 3.
The time-complexity was estimated by simulations,
it depends only linearly in the number of cells.

There are several topics for future work, like
\begin{itemize}
	\item Apply cyclic boundary conditions. 
	\item What are the challenges and new insights if we use a 3D space? 
	\item How can certain pattern characteristics 
	like the number of loops,
	the loops length, or the density be
	maximized or minimized?
	\item How can the time-complexity be computed theoretically?
	\item
	Touching points can be defined in a different way with another set of tiles.
	\item
	The generation of Hamiltonian Cycles 
  \cite{1991SpaceFilling,1992-Kwong-Hamiltonian,1997-Umans-Hamiltonian,2012Hamiltonian,CrossleyHamiltonian}
  could be addressed.
\end{itemize}



\newpage
\footnotesize
\begin{thebibliography}{00}

\bibitem{2014-Hoffmann-Agent-PathPattern}
\reff{2014-Hoffmann-Agent-PathPattern}
Hoffmann, R. 
-- How agents can form a specific pattern. 
In Cellular Automata: 11th Intern. Conf. on Cellular Automata, ACRI 2014, Proceedings 11 (pp. 660-669). Springer.
(2014)

\bibitem{2022-2019-Arxiv-Forming-Point-Patterns-by-a-Probabilistic-Cellular-Automata-Rule}
\reff{2022-2019-Arxiv-Forming-Point-Patterns-by-a-Probabilistic-Cellular-Automata-Rule}
Hoffmann, R.
-- Forming Point Patterns by a Probabilistic Cellular Automata Rule.
{\it arXiv} 2022, arXiv:2202.06656. Based on the \textit{Presentation at Summer Solstice Conference on Complex Systems (June 2019)} 

\bibitem{2021-Hoffmann-DD-FS-DominoSquareDiamond}
\reff{2021-Hoffmann-DD-FS-DominoSquareDiamond}
Hoffmann, R., Désérable, D.,  Seredyński, F.  
-- A cellular automata rule placing a maximal number of dominoes in the
 square and diamond. The Journal of Supercomputing, 77, 9069-9087.
(2021)

\bibitem{Hoffmann:Deserable-Seredynski-pact-2021b-Minimal-Covering-of-the-Space-by-Domino-Tiles}
\reff{Hoffmann:Deserable-Seredynski-pact-2021b-Minimal-Covering-of-the-Space-by-Domino-Tiles}
Hoffmann, R.; D\'{e}s\'{e}rable, D.; Seredy\'{n}ski, F.
-- Minimal Covering of the Space by Domino Tiles. In
\textit{{Parallel Computing Technologies}}; Malyshkin, V., Ed.; PaCT 2021; LNCS 12942;  {Springer: Cham, Switzerland}, 2021;
pp. 453--465. 

\bibitem{2023-Hoffmann-Loop}
\reff{2023-Hoffmann-Loop}
Hoffmann, R.  
-- Generating Loop Patterns with a Genetic Algorithm and a Probabilistic Cellular Automata Rule. 
Algorithms, 16(7), 352.
(2023)

\bibitem{2024-Hoffmann-Bialecki-Loop-AdvancesInCA}
\reff{2024-Hoffmann-Bialecki-Loop-AdvancesInCA}
Hoffmann, R.,
Bialecki, M.
-- Loop Patterns Formed by Cellular Automata.
(in publication) Chapter in Book: Advances in Cellular Automata, Springer 2024

\bibitem{Hoffmann:Deserable-Seredynski-2022-NatCom-Cellular-automata-rules-solving-the-wireless-sensor-network-coverage-problem}
\reff{Hoffmann:Deserable-Seredynski-2022-NatCom-Cellular-automata-rules-solving-the-wireless-sensor-network-coverage-problem}
Hoffmann, R.; D\'{e}s\'{e}rable, D.; Seredy\'{n}ski, F. 
-- Cellular automata rules solving the wireless sensor network coverage problem.
\textit{Nat. Comp.} 2022, {\it21}, 417--447.  

\bibitem{1991SpaceFilling}
\reff{1991SpaceFilling}
Prusinkiewicz, P.; Lindenmayer, A.; Fracccia, D.
-- Synthesis of space-filling curves on the square  grid.
In \textit{Fractals in the Fundamental and Applied Sciences};
Peitgen, H.-O., Henrique, J.M., Pencdo, L.F., Eds.;
Elsevier Science Publishers B.V.:  {Amsterdam, The Netherlands}, 1991.

\bibitem{1992-Kwong-Hamiltonian}
\reff{1992-Kwong-Hamiltonian}
Kwong, YH Harris. 
-- Enumeration of Hamiltonian cycles in P4 x Pn, and P5 x Pn.
Ars Combinatoria 33 87-96. (1992)

\bibitem{1997-Umans-Hamiltonian}
\reff{1997-Umans-Hamiltonian}
Umans, Chr. and Lenhart, W. 
-- Hamiltonian cycles in solid grid graphs.
Proceedings 38th Annual Symposium on Foundations of Computer Science. IEEE, 1997.
(1997)

\bibitem{2012Hamiltonian}
\reff{2012Hamiltonian}
Keshavarz-Kohjerdi, F.; Bagheri, A. 
-- Hamiltonian paths in some classes of grid graphs. 
\textit{J. Appl. Math.}
2012, {\it 2012}, 475087.

\bibitem{CrossleyHamiltonian}
\reff{CrossleyHamiltonian}
Crossley, M.
-- Hamiltonian Cycles in Two Dimensional Lattices. 
In \textit{Statistical Physics in Biology};  {Springer: Berlin/Heidelberg, Germany, 2013}.

\end{thebibliography}
\end{document}